\documentclass[9pt,a4paper]{article}
\usepackage[utf8]{inputenc}
\usepackage[T1]{fontenc}
\usepackage{amsmath,amsfonts,amssymb}
\usepackage{setspace}
 \usepackage{enumitem} 
   
\usepackage{graphicx}
\usepackage{caption}
\usepackage{diagbox} 
\usepackage{multirow} 
\usepackage{array} 
\usepackage[caption=false]{subfig}

\newcommand{\argmin}{\operatornamewithlimits{argmin}}

\begin{document}

\title{Eigen-Epistasis for detecting Gene-Gene interactions}
\date{}
\maketitle

\begin{center}
Virginie Stanislas$^1$, Cyril Dalmasso$^1$ and Christophe Ambroise$^1$\\ 
$^1$ Laboratoire de Math\'ematiques et Mod\'elisation d'Evry (LaMME), Universit\'e d'Evry Val d'Essonne, UMR CNRS 8071, ENSIIE, USC INRA
\end{center}

\bigskip

\noindent
Corresponding author: Virginie Stanislas, Laboratoire de Math\'ematiques et Mod\'elisation d'Evry, 23 bvd de France, 91 037 Évry Cedex, France. E-mail: virginie.stanislas@math.cnrs.fr

\newpage

	\section*{Abstract}
       	A large amount of research has been devoted to the detection
        and investigation of epistatic interactions in genome-wide
        association studies (GWASs). Most of the literature focuses on
        low-order interactions between single-nucleotide polymorphisms
        (SNPs) with significant main effects.

      In this paper we propose an original approach for detecting
        epistasis at the gene level, without systematically filtering
        on significant genes. We first compute interaction variables
        for each gene pair by finding its Eigen-Epistasis component,
        defined as the linear combination of Gene SNPs having the
        highest correlation with the phenotype. The selection of
        significant effects is done using a penalized regression method
        based on Group Lasso controlling the False Discovery Rate.

        The method is tested against two recent alternative proposals
        from the literature using synthetic data, and shows good
        performances in different settings. We demonstrate the power of 
        our approach by detecting new gene-gene interactions on three
        genome-wide association studies.

	\textbf{Key words} : Genome-wide association study, Gene-gene interactions, Epistasis, Group Lasso
	\newpage

\section*{Background}

Genome Wide Association Studies (GWASs) look for genetic markers
linked to a phenotype of interest. Typically, hundreds
of thousands of single nucleotide polymorphisms (SNPs) are studied for
a limited number of individuals using high-density
genotyping arrays. Usually the association between each SNP and
the phenotype is tested using single-marker
methods. Multiple markers may also be considered, but these are typically
selected using simple forward-selection methods.
GWASs are a powerful tool for investigating the genetic
architecture of complex diseases and have been successful in identifying
hundreds of variants. However, they have been able to explain only a
small proportion of the phenotypic variations expected
from classical family studies \cite{Manolio2009Finding}.
A number of explanations for this missing heritability have been put forward. For example, it has been suggested that shared environments among relatives are not adequately taken into account. Another suggestion is that much larger numbers of variants with small effects remain to be identified. Rare variants, which are difficult to find using existing
genotyping arrays \cite{Manolio2009Finding}, seem to be
important causal factors, and so do structural variations. But complex diseases may also be caused, at least in part, by complex genetic
structures with multiple interactions between markers (a phenomenon termed
\textit{epistasis}). Whereas in pedigree studies the genetic effect on phenotype is seen as
part of the additive 
genetic variance, in GWASs it is seen as an unmeasured interaction between genes \cite{Haig2011Does}. For example, Zuk et al. proposed a model that takes into account epistatic interaction in relation to Crohn's disease \cite{Zuk2012mystery}. They found that $80\%$ of the missing heritability could be due to genetic interactions.

In recent years a number of methods for
studying epistasis have been proposed and reported in various reviews
\cite{Niel2015survey, Wei2014Detecting, Steen2012Travelling}. They vary in
terms of their data analysis (genome-wide or filtering) and their statistical
methodology (Bayesian, frequentist, machine learning or data
mining). Most of them focus on single-locus interactions,
but considering interactions at the gene level can have several
advantages. First, given that genes are the functional unit of the genome,
results may be more biologically interpretable. Second, genetic
effects are more easily detected when SNP
effects are aggregated together. Third, gene-based analysis
simplifies the multiple testing problem by reducing the number of
variables. Several gene-gene methods have been proposed. These are based
on a summarizing step which is used to obtain information at the
gene level. In more recent methods, filters or
penalized models are used to make the method applicable
to a large number of genes, while older methods are only applicable to
two or a very limited number of genes. For the summarizing step, most
methods resort to a principal components (PC) approach, but each
method has its specific characteristics.
We describe some of these below.

Chatterjee et al. harnessed Tukey's one-degree-of-freedom method to
investigate interaction between two genes \cite{Chatterjee2006Powerful}. Their method is based on the assumption that
the SNPs included in each gene region act as surrogates for an underlying
biological phenotype. The genotypic information for the
gene region is extracted as a single component by a weighted sum of all SNPs. The weights are determined according to the SNP's correlation
with the trait. The product of the two sums is then introduced as the gene-gene interaction term into a logistic model, where marginal effects are represented by the respective sums. Building on this idea, Wang et al.
compared two different interaction tests \cite{Wang2009partial}. On the one hand, they used \textit{Principal Component
Analysis} (PCA) to summarize SNP information within a gene, and on the
other hand they used \textit{Partial Least Squares} (PLS) to extract components that summarize,
first, the information among SNPs in a gene and, second, the correlation
between SNPs and the outcome of interest. They then proposed an
interaction test based on either the first PC or
the first PLS component for each gene, and were able to show that the PCA and PLS methods often
outperformed Tukey's one-degree-of-freedom method. But it is worth noting that the main objective of these three methods was improving the detection of associations in the presence of gene-gene interactions, rather than
identifying the interactions themselves. Other approaches based on principal
component analysis have since been proposed for epistasis
detection. Li et al. proposed selecting, as the gene representation, PCs that are able to explain at least
80\% of the variation \cite{Li2009Identification}.

Genotypic data are characterized by the high correlation among markers resulting from so-called linkage disequilibrium (LD). Procedures that take LD information into account have been developed for epistasis detection.
For example, He et al. proposed an approach using LD information to weight genotype scores which are then
aggregated using principal components \cite{He2011Gene}.
Rajapakse et al. developed a gene-based test of
interactions for case-control studies which compares LD patterns
between cases and controls \cite{Rajapakse2012Multivariate}. Using the same idea, Peng et al. used a canonical correlation-based U-statistic model (CCU) to detect co-association in case-control studies \cite{Peng2010gene}.
The idea is to test for two given genes the difference between canonical correlation coefficient computed  by CCA among cases and among controls. Their work was subsequently extended to include kernel \cite{Larson2014Kernel, Yuan2012Detection}.

However most of these methods can be applied only to a reduced
number of genes. Computational constraints mean that it is not feasible to model all gene-gene interactions directly. One way of overcoming this is to
reduce the gene-gene search space by eliminating unimportant
genes, and to this end two-step procedures have been developed that first filter out specific
genes or SNPs through a genome-wide search before testing for
interactions. One example of this is the model-based kernel machine
method (3G-SPA) proposed by Li and Cui, which first performs a search for gene pairs contributing to the overall phenotypic variations \cite{Li2012Gene}. Significant pairs are then tested for interaction effects. Another attractive
alternative is offered by
penalized regression methods that select a subset of
important predictors out of a large number of potential predictors.  These
methods operate by shrinking the size of the coefficients. The
coefficients of predictors with little or no apparent effect are force to be set to zero, 
reducing the effective degrees of
freedom and in many cases making model selection possible.  A few
approaches using penalized models have been proposed. D'Angelo et al. combined principal component analysis and
lasso penalized regression \cite{DAngelo2009Combining}.
Wang et al. used a principal component analysis combined with an L1 penalty, with adaptive weights based on gene size, pathway support and effect size \cite{Wang2014Pathway}.

Here we propose a Group Lasso approach \cite{Yuan2006Model} that
takes into account the group structure of each gene in order to detect
epistasis. We introduce \textit{Gene-Gene Eigen-Epistasis} (G-GEE) as a new
approach for computing the gene-gene interaction part of the model, and we
compare G-GEE with two different interaction variable modeling
approaches inspired by previous proposals in the literature, namely PCA and PLS.  An
adaptive ridge-cleaning approach \cite{Becu2017support} is then used in order to compute p-values
for each group.

In the next section,  we detail each
model and outline the design of the simulation studies performed to
compare the performance of the different approaches. In the Results section, the findings of the simulation studies are shown, and we illustrate our approach on three real datasets
relating to ankylosing spondylitis, thyroid carcinomas and inflammatory bowel disease. The different
approaches and the results are discussed in the last section.


\section*{Methods}

We consider $n$ individuals where $\textbf{y}=(y_1,y_2,...,y_n)^T$ denotes the
vector of trait values. For each individual, genetic
variants among $G$ genes are considered.  Each gene is described by  a given number of SNPs
$p_g$ where $\sum_g p_g=p$.   The SNP matrix
$\mathbf{X}  \in \mathbb{R}^{n \times   p}$  considers an additive coding scheme in which the genotype
value of each SNP $j$ from individual $i$ is denoted  $X_{ij}\in \{1, 2, 3 \}$.
$\mathbf{X}_{i}$ is a $p$-dimensional vector of covariates for observation $i$ and for
$j \in \{1, \dots, p\}$. $\boldsymbol{X}^g$ denotes the  submatrix of $\boldsymbol{X}$ whose columns are the $p_g$ SNPs of gene $g$.
A generalized linear model is generally  assumed for GWAS, where the phenotype is
considered as a random variable $y_i$ whose conditional expectation
can be written as a function of the covariates $\boldsymbol{X}_{i}$
and their interactions $\boldsymbol{Z}_i$,

$$
g(E [y_i | X ]) =  \boldsymbol{X}_{i}^T \boldsymbol{\beta} + \boldsymbol{Z}_i^T \boldsymbol{\gamma},
$$

where
$$
\boldsymbol{\beta} =  \left( \underbrace{ \beta_{1,1}, \beta_{1,2}, ... , \beta_{1,p_1}}_{gene_1}, ... , \underbrace{\beta_{G,1}, ... , \beta_{G,p_G}}_{gene_G} \right)^T,
$$

and  $\boldsymbol{Z}_i$ is the $i$th line of the matrix of interactions and $\boldsymbol{\gamma}$ a parameter vector of appropriate dimension.  When the phenotype is binary (case control study), it is usual to assume a logistic model where $g()$ is the logit and $Y$ is assumed to follow a binomial distribution. Below we will consider only quantitative phenotypes using a classical linear model. In this case $g()$ is the identity and the residuals are assumed to be Gaussian.

The main effect of each gene is modeled through the sum of the effects of all its
SNPs. Concerning interaction effects, we compute  new
variables representing interactions between two specific genes and define
as a group all the interaction variables related to  a given pair of genes.  The matrix of interaction  is thus structured
into $G(G-1)/2$ submatrices:

$$
\boldsymbol{Z}= [ \boldsymbol{Z}^{11} \cdots  \boldsymbol{Z}^{rs} \cdots  \boldsymbol{Z}^{G(G-1)/2}]
$$

where $\boldsymbol{Z}^{rs}$  describes the interactions between the two genes $r$ and $s$. The parameter vector
  $\boldsymbol{\gamma}$ is accordingly  structured into  sub-vectors  $\boldsymbol{\gamma}^{rs}$.
We will now  present  and compare three different approaches for modeling gene-gene
interactions.

\subsection*{Modeling Gene-Gene interactions}

Let us consider two genes $r$  and $s$ described respectively by $p_r$ and $p_s$ SNPs.
A possible interaction term describing the epistasis between the two genes is

\begin{equation}
\label{eq:interation-general-form}
 {\boldsymbol{Z}_i^{rs}}^T \boldsymbol{\gamma}^{rs} =  \sum_{j=1}^{p_r} \sum_{k=1}^{p_s} \gamma_{jk}^{rs} X_{ij}^r X_{ik}^s.
\end{equation}

We hereafter set $\boldsymbol{W}^{rs}=\{ X_{ij}^r X_{ik}^s \}_{i=1\cdots n}^{j=1,\cdots,p_r;
  k=1,\cdots,p_s}$.
In this case the submatrix of interactions is
$\boldsymbol{Z}^{rs}=\boldsymbol{W}^{rs}$ and $\boldsymbol{\gamma}^{rs}=\{\gamma_{jk}^{rs}\}$ is a vector of size
$p_r p_s$.  The number of parameters in such a model is obviously too
large to be reliably estimated.  For this reason a number of papers in the literature
consider reducing the dimension of $\boldsymbol{\gamma}$.

In this paper we will consider three different methods for reducing the dimension
reduction, namely Principal Component Analysis (PCA),
Partial Least Squares (PLS), and
our proposed Gene-Gene Eigen-Epistasis approach that we have termed G-GEE.

\subsubsection*{Principal Component Analysis}

Principal Component Analysis (PCA) can reduce the number of variables describing each gene $r$ from $p_r$ to $q_r<p_r$.
Considering gene $r$ described by $p_r$ SNPs, we compute the matrix of the first $q$ principal  components

$$
C^r =  \boldsymbol{X}^{r} U^r,
$$
where $U^r$ is the matrix of the first $q_r$ principal axes. Using
$C^r$ and $C^s$ instead of $\boldsymbol{X}^{r}$ and
$\boldsymbol{X}^{s}$ in the computation of the interaction allows
the number of parameters relative to each interaction to be controlled.  This
control is achieved by choosing the number of principal components
$q$.  The PCA model that we describe draws upon ideas in
\cite{Zhang2008approach}.
The interaction term takes the form
$$
 {\boldsymbol{Z}_i^{rs}}^T \boldsymbol{\gamma}^{rs} = \sum_{j=1}^q \sum_{k=1}^q \gamma_{jk}^{rs} C_{ij}^r C_{ik}^s.
$$

Relating this expression to the general form of the interaction term $\boldsymbol{W}_i^{rs}$ described above,
we can see that performing PCA prior to computing the interactions is
a means of constraining the linear interaction term of Equation \ref{eq:interation-general-form}.

The submatrix of interactions is
$\boldsymbol{Z}^{rs}=\{ C_{ij}^r C_{ik}^s \}_{i=1\cdots n}^{j=1, \cdots, q;k=1, \cdots, q }$, and $\boldsymbol{\gamma}^{rs}=\{\gamma^{rs}_{jk}\}$ is a vector of size
$q^2$  describing the interaction between genes
$r$ and $s$. In particular, if a single principal component is chosen, there will be only one parameter to estimate per interaction.

\subsubsection*{Partial Least Squares}
Wang et al. proposed an alternative  method for integrating interactions
using a PLS approach \cite{Wang2009partial}. Let
$(\boldsymbol{X}^{r},\boldsymbol{X}^{{s}})$ be the genotypic matrix
for the given pair of genes $(r,s)$. Their approach computes 
the components that maximize
$cov^2(\boldsymbol{X}^r \boldsymbol{u}, \boldsymbol{T}\boldsymbol{v})$, with
$\boldsymbol{T}=(\boldsymbol{y},\boldsymbol{X}^{{s}})$ and
$(\boldsymbol{u},\boldsymbol{v})$ the weight vectors.
The interaction of a couple of genes (r, s) is then represented by the first q components:

$$
 {\boldsymbol{Z}_i^{rs}}^T \boldsymbol{\gamma}^{rs} =\sum^q_{j=1} \gamma^{rs}_j T^{rs}_{ij}.
$$

In this approach phenotypic information is retained when the
interaction variables are constructed.

\subsubsection*{Gene-Gene Eigen-Epistasis}
We propose an original approach for modeling interactions. The
general idea is to consider the interaction variable between the two genes $r$ and $s$
as a function $f_{\boldsymbol{u}}(\boldsymbol{X}^r,\boldsymbol{X}^s)$
parameterized by $\boldsymbol{u}$.  One way to estimate
$\boldsymbol{u}$ is to maximize the correlation between the
interaction function  and the phenotype:

$$
\hat{\boldsymbol{u}}= arg \max_{\boldsymbol{u}, \| \boldsymbol{u}\|=1 } cov^2(\boldsymbol{y},f_{\boldsymbol{u}}(\boldsymbol{X}^r,\boldsymbol{X}^s)).
$$
If we consider the function $f$ to be linear, our problem becomes easily tractable and has only one solution. Setting
$$
 \boldsymbol{Z}^{rs}=f_{\boldsymbol{u}}(\boldsymbol{X}^r,\boldsymbol{X}^s) = \boldsymbol{W}^{rs}\boldsymbol{u},
$$
where  $\boldsymbol{W}^{rs}=\{ X_{ij}^r X_{ik}^s \}_{i=1\cdots n}^{j=1 \cdots , p_r; k=1,\cdots, p_s}$ and $\boldsymbol{u} \in \mathbb{R}^{p_r p_s}$ we obtain the following problem:
\begin{equation}
\max_{\boldsymbol{u},\|\boldsymbol{u}\|=1}||\hat{cov}[{\boldsymbol{W}^{rs}}\boldsymbol{u},\boldsymbol{y}]||^2=
\max_{\boldsymbol{u},\| \boldsymbol{u}\|=1}||\boldsymbol{u}^T {\boldsymbol{W}^{rs}}^T \boldsymbol{y}||^2
=\max_{\boldsymbol{u},\| \boldsymbol{u}\|=1}\boldsymbol{u}^T {\boldsymbol{W}^{rs}}^T\boldsymbol{y}\boldsymbol{y}^T \boldsymbol{W}^{rs} \boldsymbol{u} \quad .
\end{equation}

The solution $\boldsymbol{u}$ is the eigenvector corresponding to the largest eigenvalue of the matrix ${\boldsymbol{W}^{rs}}^T\boldsymbol{y}\boldsymbol{y}^T \boldsymbol{W}^{rs}$, which is the vector ${\boldsymbol{W}^{rs}}^T\boldsymbol{y}$.
The complexity of  computing  $\boldsymbol{u}$ is therefore in  $O(n p_r p_s)$.
We then use the projection of the  matrix $\boldsymbol{W}^{rs}$ on $\boldsymbol{u}$ as the interaction variable. The resulting Eigen-Epistasis vector $\boldsymbol{Z}$ is the linear combination of all the SNP-SNP interactions being 
the most correlated with the phenotype.
In its construction, G-GEE has similarities with PLS. The main difference lies in the original design matrix. PLS searches for components that maximize $ cov^2(\boldsymbol{X}^r \boldsymbol{u}, \boldsymbol{y} \boldsymbol{X}^sv)$, whereas G-GEE retains the component that maximizes $cov^2(\boldsymbol{y},\boldsymbol{W}^{rs}\boldsymbol{u})$, with $\boldsymbol{W}^{rs}$ the matrix of all pairwise interaction between the two genes $r$ and $s$.
Like PLS, G-GEE takes phenotypic information into account in the construction of the interaction variables.
Other methods as such as CCU \cite{Peng2010gene} and the kernel versions of CCU \cite{Larson2014Kernel, Yuan2012Detection} that we referred to in the introduction also consider the phenotype in their construction, but these methods can be applied only to case-control problems.

\subsection*{Estimation of coefficients}

We propose a Group Lasso approach \cite{Yuan2006Model} for estimating the parameters of linear or logistic (case control) regression.
A group comprises either the SNPs of a given gene, or interaction
terms relative to a given gene-pair interaction. In the particular case of linear regression, the model parameters are estimated by:

\begin{equation*}
 \hat{\boldsymbol{\theta}}=(\hat{\boldsymbol{\beta}},\hat{\boldsymbol{\gamma}}) = \argmin_{\boldsymbol{\beta},\boldsymbol{\gamma}}  {\left(\sum_i{(y_i - \boldsymbol{X}_i \boldsymbol{\beta} - \boldsymbol{Z}_i \boldsymbol{\gamma})^2} +\lambda \left[
\sum_{g} \sqrt{p_g}||\boldsymbol{\beta}^g||_2 + \sum_{rs} \sqrt{p_r p_s}||\boldsymbol{\gamma}^{rs}||_2\right] \right)} \quad ,
\end{equation*}

The parameter $\lambda$ is selected by cross-validation.

In order to improve estimation accuracy and to obtain p-values for
each of the selected groups, we use the adaptive ridge cleaning approach
proposed by B\'ecu et al. \cite{Becu2017support}. This screen and clean procedure
is a two-stage method. The group lasso model is first fitted on half
of the data. The coefficient of the candidate groups selected by the
model are then introduced into a ridge regression model fitted on the
second half of the data with a specific penalty that allows
the group structure to be taken into account. For each group the significance of the regression
coefficients is estimated using permutation tests.

\subsection*{Simulation Design}

To evaluate the performance of the proposed approach, we conducted
two simulation studies, the first using simulated data and the second using a real dataset relating to ankylosing spondylitis. In each case we compared the proposed G-GEE model to the two
other interaction variable modeling approaches.  
The first simulation corresponds to a simplified context where all parameters were controlled and external interference limited, while the second simulation corresponds to a realistic context with a realistic pattern of MAF and LD.

\subsubsection*{Design \label{sec:design}}

\paragraph*{Genotypes}

Our first (simplified) simulation study was adapted from the model used in \cite{Wu2009Genome} with an extension to control the minor allele frequency (MAF) of each SNP.
The $n$ lines of the  genotype matrix are an i.i.d. sample from  a multivariate random vector
$\boldsymbol{X}_i\sim \mathcal{N}_p
(\boldsymbol{0},\boldsymbol{\Sigma})$. The correlation matrix
$\boldsymbol{\Sigma}$ is block diagonal, each block corresponding to a
gene. Two variables belonging to the same gene are correlated at level
$\rho=0.8$ while all other correlations are null.  Each SNP (column of
the genotype matrix) is  randomly assigned an MAF
$p$ from a uniform distribution between 0.05 and 0.5.  An MAF value of
$0.2$ is assigned to all causal SNPs.
The genotype frequencies derived from the Hardy-Weinberg equation are
then used to discretize  $X_{ik}$ values to $1$, $2$ or $3$. In practice,
$X_{ik}$ is set to $1$  if  $X_{ik}<q_{p^2;N(0,1)}$, $X_{ik}$ is set to $3$ if
$X_{ik}<q_{(1-p)^2;N(0,1)}$ and $X_{ik}$ is set to $2$  otherwise.

In the second (realistic) simulation study using a real ankylosing spondylitis dataset, genes are randomly selected. The number of SNPs composing each genes varies according to the selection.

\paragraph*{Phenotypes}

For both simulation studies, we generated phenotype vectors using two different schemes. Our first scheme corresponds to the model proposed by Wang et al. \cite{Wang2014Pathway} (which, for the sake of brevity, we will refer to hereafter as the ``Wang Pathway'' model):

\begin{equation}
Y_i = \beta_0 + \sum_{g} \beta_{g}   \left( \sum_{k  \in \mathcal{C}}  X_{ik}^g  \right)
+ \sum_{rs}   \gamma_{rs}  \left(  \sum_{(j,k) \in \mathcal{C}^2 } X_{ij}^r X_{ik}^s \right)  + \epsilon_i ,
\end{equation}

where $\mathcal{C}$ and $\mathcal{C}^2$ are respectively the set of causal SNPs and causal interactions, and $\epsilon_i$ a random Gaussian variable.  For each causal gene $g$,  we consider two  causal SNPs and a
coefficient $\beta_{g}$ is  assigned to the standardized sum of these
causal SNPs. In the same way, for the interactions, all the causal SNPs from a causal pair $(r,s)$ are  pairwise multiplied and a coefficient $\gamma_{rs}$ is assigned to the standardized sum of the product.

Our second scheme for simulating phenotypes is based on the following model:

\begin{equation}
Y_i = \beta_0 + \sum_{g} \beta_{g}    \left( \sum_{k  \in \mathcal{C}}  X_{ik}^g \right)
+ \sum_{rs}   \gamma_{rs}  \left( \sum_{(j,k) \in \mathcal{C}^2 } C_{ij}^r C_{ik}^s  \right)  + \epsilon_i .
\end{equation}

The difference with the first model concerns the simulation of the interaction effect. In the second model  the interaction effect for a causal couple $(r,s)$ is defined as the product of the first PCA component $\boldsymbol{C}_{.1}^r$ of gene $r$ and the first PCA component $\boldsymbol{C}_{.1}^s$ of gene $s$.
	
In both models, $\beta_0$ is set to $0$, and $\epsilon_i$ are  generated independently from a $\mathcal{N}(0,\sigma^2)$, with $\sigma^2$ determined from  the coefficient of determination $R^2$ that calibrates the strength of the association. Both simulation models can be written as
$y_i = \boldsymbol{X}_i^T \boldsymbol{\beta} + \boldsymbol{Z}_i^T
\boldsymbol{\gamma} + \epsilon_i$ where  $\boldsymbol{X}$ the marginal
effect genotype matrix and  $\boldsymbol{Z}$ the interaction effect matrix.

Let us denote
$\boldsymbol{Q} \boldsymbol{\phi}= [\boldsymbol{X}, \boldsymbol{Z}]
 \begin{bmatrix}
   \boldsymbol{\beta} \\
   \boldsymbol{\gamma}
\end{bmatrix}$
and

\begin{align*}
R^2&=\dfrac{\sum(\boldsymbol{Q_i\phi} - \bar{y})^2}{\sum(\boldsymbol{Q_i\phi} + \epsilon_i - \bar{y})^2}\\
&=\dfrac{\sum(\boldsymbol{Q_i\phi} - \bar{y})^2}{\sum(\boldsymbol{Q_i\phi} - \bar{y})^2 + \sum\epsilon_i^2 + \sum2(\epsilon_i(\boldsymbol{Q_i\phi}-\bar{y}))}\\
&=\dfrac{\sum(\boldsymbol{Q_i\phi} - \bar{y})^2}{\sum(\boldsymbol{Q_i\phi} - \bar{y})^2 + \textrm{n }\hat{\textrm{var}}(\epsilon_i) + 2 \textrm{n } \hat{\textrm{cov}}(\epsilon_i,\boldsymbol{Q_i\phi}-\bar{y})}.
\end{align*}

We remark that:

\begin{align*}
 2 \textrm{n } {\textrm{cov}}(\epsilon_i,\boldsymbol{Q_i\phi}-\bar{y}) &= 2\textrm{n cov}(\epsilon_i,\boldsymbol{Q_i\phi}-\dfrac{\sum_j y_j}{n})\\
 &=2\textrm{n cov}(\epsilon_i,\boldsymbol{Q_i\phi})-\sum_j \dfrac{2\textrm{n}}{\textrm{n}}\textrm{cov}(\epsilon_i,y_j)\\
 &=0 - 2\textrm{cov}(\epsilon_i,\epsilon_i)=-2\sigma^2
 \end{align*}

Thus, replacing $\hat{\textrm{var}}(\epsilon_i)$  by $\sigma^2$, and
$\hat{\textrm{cov}}(\epsilon_i,\boldsymbol{Q_i\phi}-\bar{y})$ by $-\sigma^2/n$, we obtain
$R^2 \approx \dfrac{\sum(\boldsymbol{Q_i\phi} - \bar{y})^2}{\sum(\boldsymbol{Q_i\phi} - \bar{y})^2 + \textrm{n}\sigma^2 -2\sigma^2}$. This relation between $R^2$ and $\sigma^2$
gives us an expression for $\sigma^2$ that depends on $R^2$,
$\sigma^2 = \dfrac{(R^2 - 1) \sum(\boldsymbol{Q_i\phi} - \bar{y})^2}{R^2 (2-n)}$.

We looked at how much of the coefficient of determination $R^2$ is explained by main effects, and how much is explained by interaction effects,
in order to determine their respective roles in the model.

For a similar reason, when simulating phenotypes, Wang et al. \cite{Wang2014Pathway} examined how much of partial $R^2$ was due to interaction effects.
They selected coefficient values so that 30\% of the partial $R^2$ was explained by interaction effects. Li and Cui \cite{Li2012Gene} did not use the $R^2$ directly, but they simulated data assuming different proportions of interaction effects among the total genetic variance.
In our study, once the phenotype $y$ had been set for each simulated design matrix, we
computed how much of the $R^2$ could be attributed to interaction and main effects
as $p_I=\dfrac{R^2_I}{R^2_T}$  and $p_M=\dfrac{R^2_M}{R^2_T}$ respectively, with $R^2_I$ the R-square value
for the model containing only simulated interaction effects, $R^2_M$ the R-square value where there were only simulated main effects, and
$R^2_T$ R-square value where there were both simulated main effects and simulated interaction effects.

\paragraph*{Scenarios}

In the first (simplified) simulation study, genotypes are simulated as described in the design. 
We considered six genes, each composed of six SNPs and for 600 subjects. 
We define one causal interaction between genes and two causal genes with main effects, and the simulation takes place using two alternative simulation settings:
\begin{itemize}
\item (1) one interaction and two main effects involving the same genes
\item (2) one interaction and two different main effects
\end{itemize}
For these two settings, different coefficients of determination, from $0.05$ to $0.7$, are considered and 1000 iterations are performed.
\bigskip

In the second (realistic) simulation study, genotypes come from a real dataset comprising 763 individuals. At each iteration we randomly select six genes of various size (from 1 to 1119 with a median of 2 SNPs) in the dataset. 
 We consider the five following settings: 
\begin{itemize}
\item (1) one interaction and two main effects involving the same genes
\item (2) one interaction and two different main effects
\item (3) one interaction effect only
\item (4) two main effects only
\item (5) no effects
\end{itemize}
For each setting, coefficients of determination, from $0.1$ to $0.4$, are considered and 500 iterations are performed. 

For both simulation studies, main effects
and interaction effects are  weighted  with the same coefficient
values ($\beta_g=\gamma_{rs}=2, \forall g,r,s$). 
For each interaction, the power is estimated as the proportion of detected interactions over the total number of simulations.

\subsection*{Real data illustration}

To illustrate our approach we applied the proposed method on three real datasets related to ankylosing spondylitis, thyroid carcinomas and inflammatory bowel disease.

The dataset regarding ankylosing spondylitis consists of the French subset of the large study of the International Genetics of Ankylosing Spondylitis (IGAS) study \cite{Cortes2013Identification}. For this subset, unrelated cases were recruited through the Rheumatology clinic of Ambroise Paré Hospital (Boulogne-Billancourt, France) or through the national self-help patients' association:  "Association Française des Spondylarthritiques". Population-matched unrelated controls were obtained from the "Centre d'Etude du Polymorphisme Humain", or were recruited as healthy spouses of cases. The protocol was reviewed and approved by the Ethics committee of the Ambroise Paré hospital. All participants gave their informed consent to the study. 
The application on thyroid carcinomas was carried out on a public dataset that came from the study of Luz\'on-Toro el al. on identification of epistatic interactions in two different types of thyroid carcinomas \cite{Luzon-Toro2015Identification}. Finally, we used the Wellcome Trust Case-Control Consortium genome-wide association dataset to study Inflammatory Bowel Disease.

%

\section*{Results}

\subsection*{Simulation studies}

In the following, we will refer to the different simulation settings by using letters as described in Table~\ref{tab: set}.

\begin{table}[h!]
\caption{Effects simulated in each settings and referring names according to the phenotype simulation model}
\centering
\begin{tabular}{|l|c|c|>{\centering\arraybackslash}p{2.8cm}|>{\centering\arraybackslash}p{2.8cm}|}
   \hline
   \multicolumn{3}{|c|}{Settings} & \multicolumn{2}{c|}{Names }\\
\hline 
id & Main effects & Interaction effects & Wang Pathway model & PCA model\\
\hline 
1 & Genes 1 \& 2  & Genes 1 x 2 & A & B \\
\hline
2 & Genes 1 \& 2  & Genes 3 x 4 & C & D \\
\hline
3 & - & Genes 1 x 2 & E & F \\
\hline
4 &  Genes 1 \& 2 & - & \multicolumn{2}{c|}{OME} \\
\hline
5 &  -  & - & \multicolumn{2}{c|}{NE} \\
\hline
\end{tabular}

      \label{tab: set}
\end{table}

\subsubsection*{Results from the simplified simulation study}
Figure 1 shows results obtained for the two settings. The first column gives the estimated power to detect the gene interaction as a function of the $R^2$ values.
 The last two columns show heatmap
matrices reflecting the proportion of significant values for each variable
and each method over the 1000 simulations for different $R^2$ values.

\begin{figure}[h!]
\includegraphics[width=12.5cm]{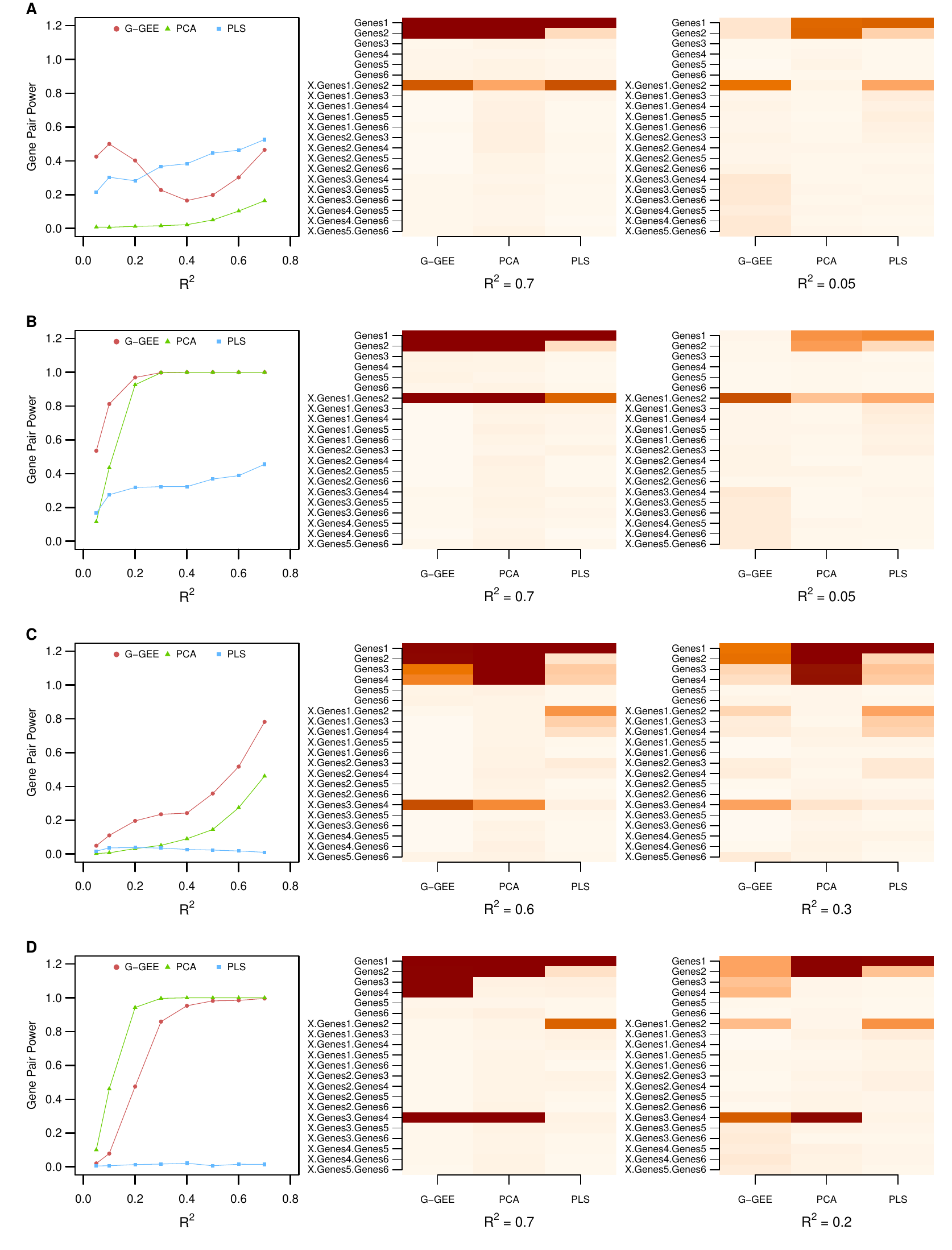}
            \caption{\textbf{Power and discoveries under a simplified context}
       The figures in the first column shows the power to detect interaction effects of the three methods depending on the $R^2$. The last two columns show the ratio of the number of times where each variable was significant to the total number of simulations for a given $R^2$. The panels \textbf{A}, \textbf{B}, \textbf{C} and \textbf{D} refer to the different simulation settings described in Table 1.}

         \label{fig: cadre}
  \end{figure}

In the first setting (Figure~\ref{fig: cadre}(A,B)), we consider genes 1 and 2, both having main and interaction effects. When the phenotype is simulated using the Wang Pathway model, the G-GEE and PLS
methods have a higher power to detect the interaction effect than PCA method, which tends to identify only the two
main effects of the two genes (Figure~\ref{fig: cadre}(A)).
Whereas for PCA and PLS the power is nondecreasing with $R^2$, for G-GEE we observe a
U-shaped curve. For the smallest $R^2$ values, which correspond
to the most difficult cases, the power of G-GEE to detect the
interaction tends to decrease. When $R^2$ values
reach $0.4$, G-GEE's power to detect the
interaction starts to increase. The situation is different
for the main effects, since G-GEE's power
to detect these increases continuously with $R^2$ [see Additional file 1].
For PLS, the power to detect the
interaction effect is continuously nondecreasing. Note, however,
that for this method one of the two main effects
(here gene 1) is detected to the detriment of the second, regardless of the value of $R^2$.
In the PCA phenotype simulation model  (Figure~\ref{fig: cadre}(B)), G-GEE has a higher
power than the other methods to detect interaction effects while retaining a good specificity, whatever
the value of $R^2$. The reasonably high power of the PCA
method can be explained by the similarity between the phenotype simulation model and the estimation model.
It is worth noting that in this first setting, only a few variables are falsely significant, which
reflects a good specificity for all methods (the worst being for
the gene 3 $\times$ gene 4 interaction variable in the case of the Wang Pathway model and $r^2=0.1$, where the false discovery rate is  $0.068$).

In the second setting (Figure~\ref{fig: cadre}(C,D)), genes 1 and 2 have only main effects, and genes 3 and 4 have only an interaction effect. When the phenotype is simulated using the Wang Pathway model, the interaction power of G-GEE is uniformly higher than that of the other methods (Figure~\ref{fig: cadre}(C)).
For all values of $R^2$, PCA tends to detect false main
effects for genes 3 and 4, but not to detect interaction effects.
In the PCA phenotype simulation (Figure~\ref{fig: cadre}(D)), PCA has a good power to detect interaction effects, but once again these good performances can be explained by the similarity between the simulation model
and the estimation model. The interaction power for G-GEE is
lower, but still good.  With this model,
only G-GEE tends to attribute a false main effect to genes 3 and 4.
In this second setting, whatever the phenotype simulation model, the power of the PLS method is almost
null. PLS identifies only the first gene as having a main effect,
while the effects of genes 3 and 4 are not detected, whether as
main or as interaction effects. Moreover, PLS tends to attribute a false interaction effect between genes 1 and 2.

To evaluate the performances of the different methods in a more complex context, we also consider a setting where we simulate 25 genes with four causal
interactions between genes, and two genes with causal main effects. In these simulations, interaction genes are different from main effect genes, and we only consider the case where $R^2=0.7$. The results of this setting reflect the good performance of the G-GEE method over PCA and PLS in detecting interaction in a context where further interactions and different main effects are simulated  [see Additional file 2].

\subsubsection*{Results from the realistic simulation study}
\begin{figure}[h!]
\includegraphics[width=12.5cm]{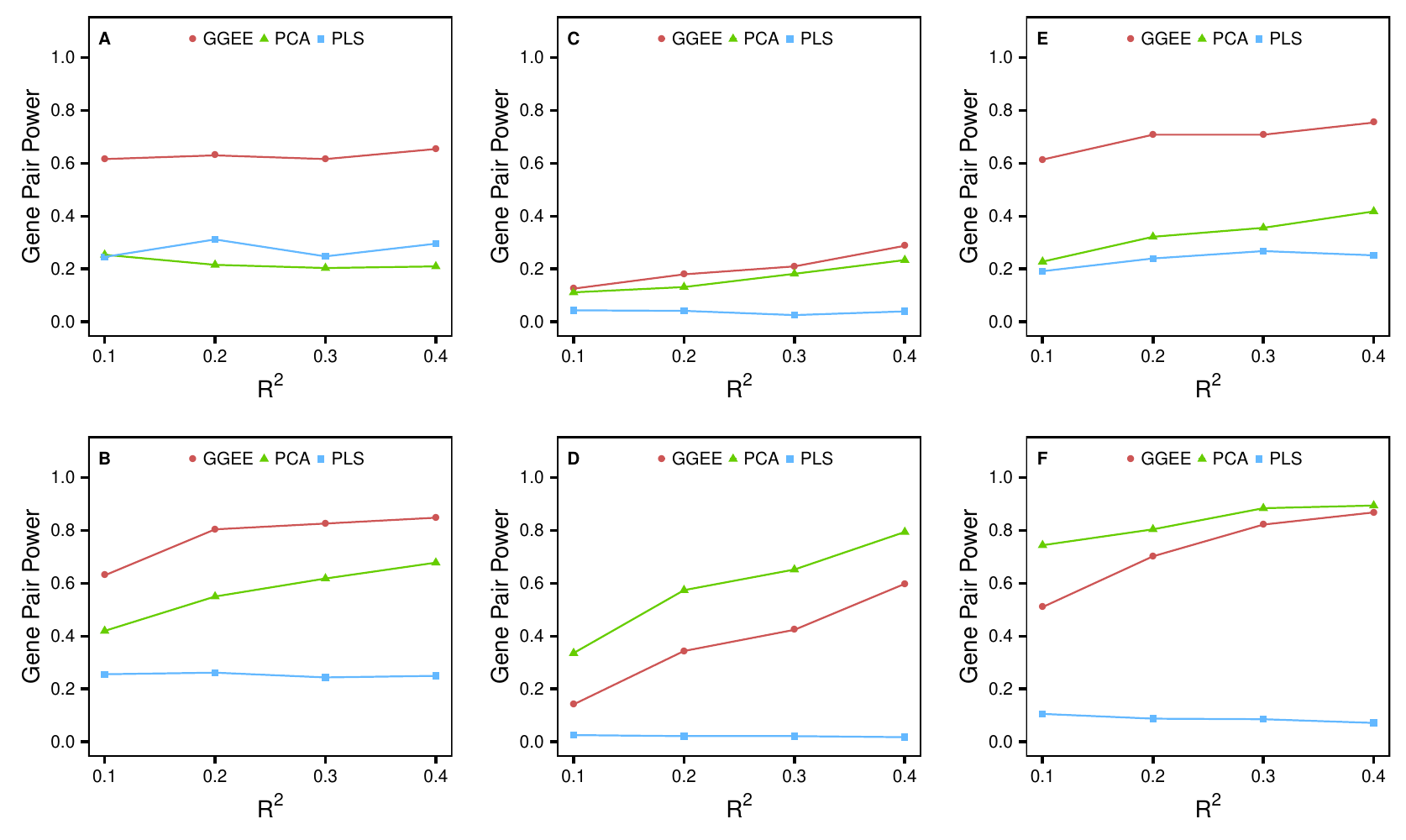}
            \caption{\textbf{Power under a realistic context}
             The figures show the power to detect interaction effects of the three methods depending on $R^2$.  The panels \textbf{A}, \textbf{B}, \textbf{C}, \textbf{D}, \textbf{E} and \textbf{F} refer to the different simulation settings described in Table 1.}

             \label{fig: Compa}
  \end{figure}

\begin{figure}[h!]
\includegraphics[width=12.5cm]{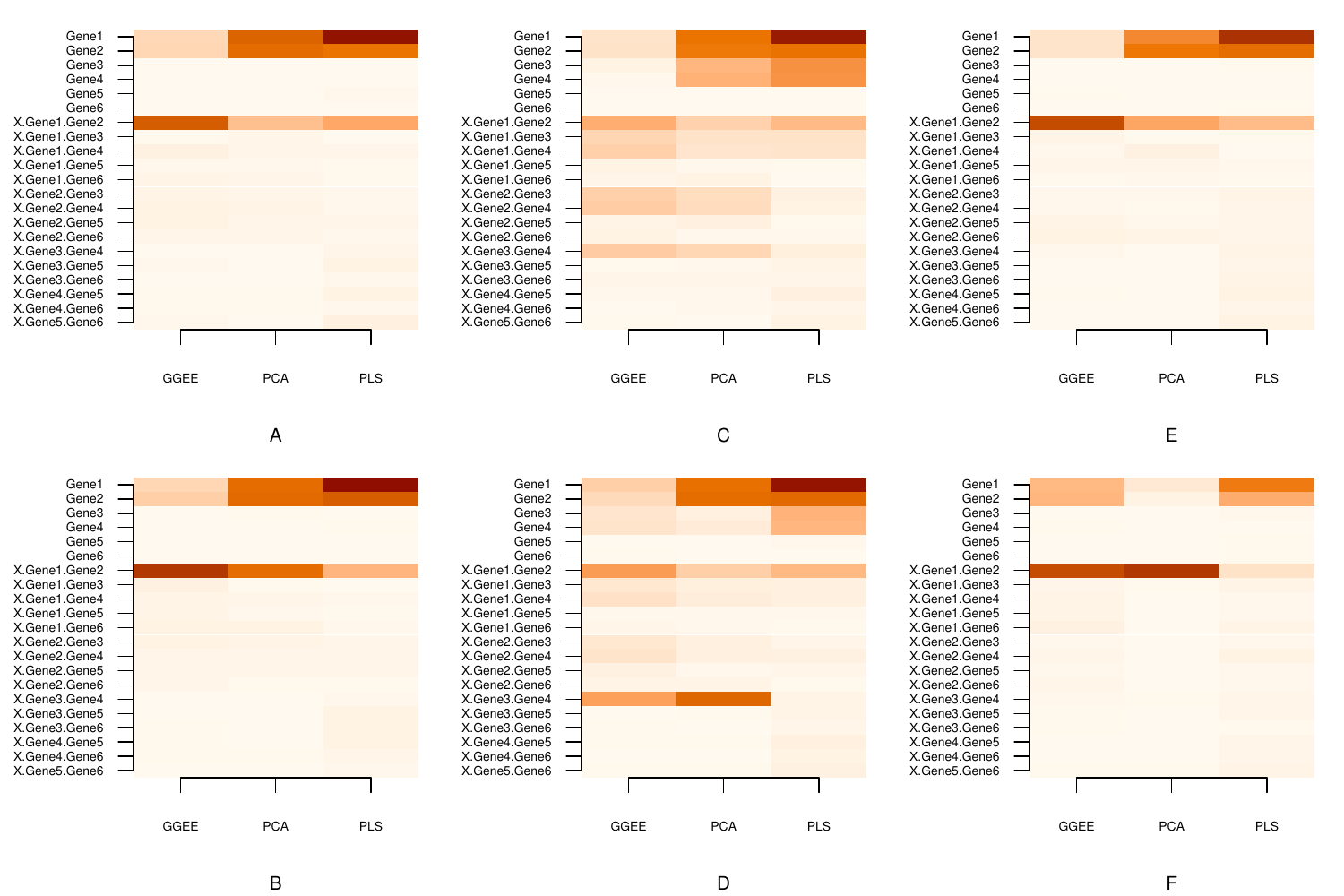}
        \caption{\textbf{Discoveries under a realistic context}
          Heatmaps of the ratio of the number of times where each variable was significant to the total number of simulations for $R^2=0.2$.The panels \textbf{A}, \textbf{B}, \textbf{C}, \textbf{D}, \textbf{E} and \textbf{F} refer to the different simulation settings described in Table 1.}
             \label{fig: MAT}
      \end{figure}

Figures~\ref{fig: Compa} and~\ref{fig: MAT} show results for the first three settings. Figure~\ref{fig: Compa} shows the power to detect gene interaction depending on the $R^2$, and Figure~\ref{fig: MAT} shows heatmaps of significant effects when  $R^2=0.2$. In both figures, the upper row relates to phenotypes simulated using the Wang Pathway model, and the lower row to phenotypes simulated using the PCA model.

In the first setting  (Figure~\ref{fig: Compa} (A, B), Figure~\ref{fig: MAT} (A, B)), the same two genes (genes 1 and 2) are simulated with main and interaction effects. In this setting G-GEE has the best power to detect interaction effects for all  $R^2$ values. The interaction power of PLS remains close to $0.3$. The power of PCA depends on the phenotype simulation model. When the phenotype is simulated using the Wang Pathway model, the power is similar to PLS. When it is simulated using PCA it increases continuously, because of the similarity between the phenotype simulation model and the estimation model. Looking at the heatmaps (Figure~\ref{fig: MAT} (A, B)) we can see that only a few variables are falsely significant. We also observe that unlike G-GEE, PCA and PLS can detect the two simulated main effects, with a preference for gene 1 in the case of PLS. These results are obtained when $R^2=0.2$, but other $R^2$ values give similar results [see Additional file 3].

In the second setting (Figure~\ref{fig: Compa} (C, D), Figure~\ref{fig: MAT} (C, D)), main effects are simulated for genes 1 and 2, and one interaction is simulated between genes 3 and 4. In this setting the power of PLS to detect interaction effects is almost null, while the respective powers of PCA and G-GEE are different, according to which phenotype simulation model is used. Both methods have a higher power when the phenotype is simulated using the PCA model. Regarding the detection of main effects, the results are similar to the first setting, with G-GEE less successful than PCA and PLS (Figure~\ref{fig: MAT} (C, D)). But unlike in the first setting, here some variables are falsely significant. False detections among interaction variables are more pronounced for G-GEE and concern genes that have been simulated to have only main effects. False detections among main effects are more pronounced for PCA and PLS when the phenotype is simulated using the Wang Pathway model and concern genes that have been simulated to have an interaction effect. Under the PCA phenotype model, false detections among main effects are more pronounced for PLS and G-GEE when $R^2$ values are higher [see Additional file 3].

In the third setting, where only one interaction is simulated between genes 1 and 2, G-GEE has a higher power to detect interaction than PLS and PCA  when the phenotype is simulated using the Wang Pathway model (Figure~\ref{fig: Compa} E). The power of PCA is higher in the PCA phenotype simulation model because of its similarity to the estimation model, whereas the power of PLS is almost null (Figure~\ref{fig: Compa} F). In the Wang Pathway phenotype simulation model, PCA and PLS both falsely detect main effects. In the PCA phenotype simulation model, the false detections are made by PLS and G-GEE (Figure~\ref{fig: MAT} (E, F)). In all cases these false detections concern genes that are simulated to have an interaction effect.

\begin{figure}[h!]
\includegraphics[width=12.5cm]{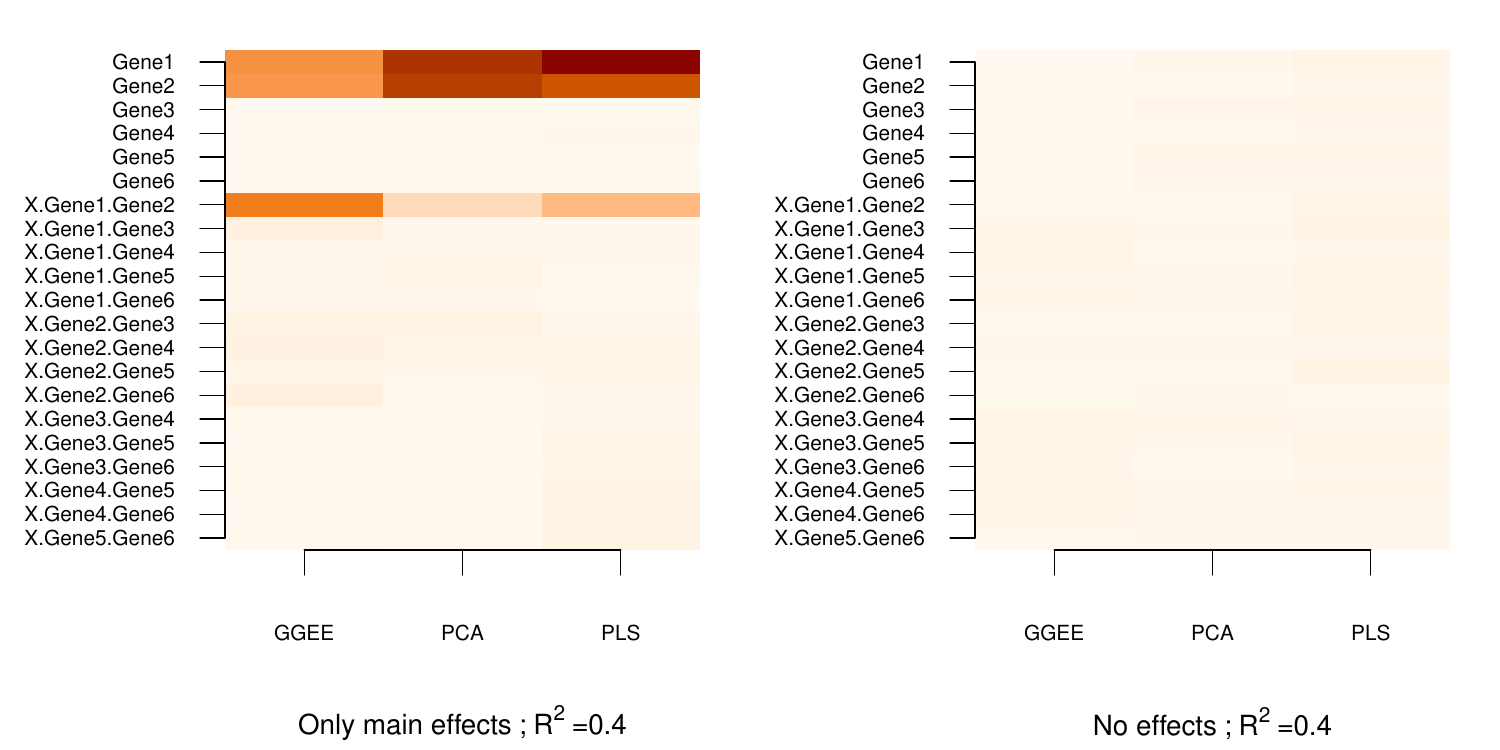}
        \caption{\textbf{Discoveries for the fifth and the sixth settings}
       Heatmaps of the ratio of the number of times where each variable was significant to the total number of simulations for $R^2=0.4$  when only main effects are simulated for gene 1 and  gene 2 (left), and when no effects are simulated (right).}
           \label{fig: OnlyMain-none}
      \end{figure}

Figure~\ref{fig: OnlyMain-none} shows the results for the fourth and fifth settings. The heatmap on the left corresponds to the fourth setting, where only two main effects are simulated. We remark that all methods successfully identify the main effect, PCA and PLS doing so with a higher power. False detections corresponding to the respective interaction effects are observed for G-GEE, and to a lesser extent for PLS. The figure on the right corresponds to the fifth setting, where no specific effects are simulated and the result shows that all three methods perform well with very few false detections.

In all settings, estimating the coefficients with the group lasso is more computationally expensive than constructing the interaction variables. G-GEE and PCA are quite similar in terms of computation time, whereas in some settings PLS has a slightly greater execution time than other methods. Note that the time required by G-GEE for constructing the interaction variables varies according to the number of SNPs that constitute each gene [see Additional file 4].

\subsubsection*{Percentage of $R^2$ attributable to interaction and main effects respectively}

\begin{table}[h!]
\caption{Average percentage of $R^2$ attributable to interaction and main effects, by setting, in the first simulation study.}
\centering
{\footnotesize
\renewcommand{\arraystretch}{1.2}
\begin{tabular}{|l|r|r|r|r|r|r|r|r|}
   \hline
 & \multicolumn{2}{c|}{A} & \multicolumn{2}{c|}{B} & \multicolumn{2}{c|}{C} & \multicolumn{2}{c|}{D}  \\
\hline
  & $R^2$=0.7 & $R^2$=0.05 &  $R^2$=0.7 &  $R^2$=0.05 &  $R^2$=0.6 &  $R^2$=0.3 & $R^2$=0.7 &  $R^2$=0.2 \\
  \cline{2-3}
\hline
$p_I$ & 97.73 & 92.08 & 33.11 & 32.80 & 33.32 & 33.47 & 33.51 & 33.57 \\
\hline
$p_M$ & 98.84 & 95.57 & 66.42 & 66.97 & 66.60 & 66.57 & 66.70 & 66.56 \\
\hline
\end{tabular}
}
      \label{tab: R2}
\end{table}

\begin{table}
  \caption{Average percentage of $R^2$ attributable to interaction and main effects, by setting, in the second simulation study.}
\centering
{\footnotesize
\renewcommand{\arraystretch}{1.2}
\begin{tabular}{|l|r|r|r|r|r|r|r|}
\hline
  & A $R^2$=0.2 & B $R^2$=0.2 & C $R^2$=0.2 & D $R^2$=0.2 & E $R^2$=0.2 & F $R^2$=0.2 & OME $R^2$=0.4 \\
\hline
$p_I$& 94.01 & 52.04 & 33.36 & 33.27 & 100 & 100 & 0 \\
\hline
$p_M$ & 99.08 & 78.62 & 66.60 & 66.87 & 0 & 0 & 100 \\
\hline
\end{tabular}
}
      \label{tab: R2RD}
\end{table}

Using each setting in both simulation studies, we determine the $p_I$ and  $p_M$ average
values that correspond to the proportion of the $R^2$ attributed to
interaction and main effects, respectively.
For most settings, the  $p_I$  depends on the number of simulated
effects. With one interaction and two main effects simulated the $R^2$
part attributable to interaction effects is around 33\%
(Table~\ref{tab: R2} (B, C, D), Table~\ref{tab: R2RD} (C, D)). For the setting with numerous effects [see Additional file 2], the
average $p_I$ is 67\% because we consider four interaction effects for only two main effects. Finally, as expected, when only interaction effects are simulated, the average  $p_I$ is 100\% (Table~\ref{tab: R2RD} (E, F)) and 0\% when only main effects are simulated (Table~\ref{tab: R2RD} (OME)).
However, the $R^2$ distribution between main and interaction effects is not distinguishable in the setting where the phenotype is simulated using the Wang Pathway model with the same main and interaction effects.  The $p_I$ and  $p_M$ values are all above 90\%  (Table~\ref{tab: R2} (A), Table~\ref{tab: R2RD} (A)). In the second simulation study, the $R^2$ distribution is also not well divided between main and interaction effects when the phenotype is simulated under the PCA model, though $p_M$ is still higher than $p_I$ (Table~\ref{tab: R2RD} (B)).

\subsubsection*{Conclusions}

The results obtained in both simulation studies point to a certain
confusion between main and interaction effects.
When simulated interaction and main effects involve different genes,
the methods tend to detect as interaction effects the pairs of genes simulated to have main effects and, conversely, to detect as main effects the genes simulated to having interaction effects.

Overall, G-GEE tends to detect more false interactions than false main effect whereas PLS and PCA tend to detect more false main effects though PLS tends to attribute a false interaction effect between genes 1 and 2.
This type of confusion may explain the U-shaped
power curve for G-GEE observed in the first simulation study (Figure~\ref{fig: cadre}(A)). As the problem becomes harder, the genetic effects of both genes are preferentially assigned to the interaction effect, implying a better power to detect interaction where $R^2$ values are small. Finally, we remark that for G-GEE false detections of main effects are more frequent when the PCA phenotype simulation model is used, whereas for the PLS and PCA methods, where the number of false detections for main effects is higher when the Wang Pathway phenotype simulation model is used.

Other observations regarding the power of the different methods can be made with these simulation results. PLS has more trouble than PCA and G-GEE in detecting interaction effects, and has a tendency to detect the first main effect with a higher power than the second main effect when two main effects are simulated. For all methods, the power to detect interactions increases more slowly with respect to $R^2$  when simulations are performed using real data genotypes than with fully simulated genotypes, but we observe that in the first setting the curve representing the interaction power of G-GEE is detached from the others, reflecting the superior performance of G-GEE over PLS and PCA. Note that the power of G-GEE to detect main effects is always less than that of PCA and PLS when $R^2 < 0.4$ [see Additional file 1 and 5]. In short, G-GEE performs better when detecting interactions than  when detecting main effects.

\subsection*{Real data illustrations}

\subsubsection*{Ankylosing spondylitis}
Ankylosing spondylitis (AS) is a common form of inflammatory arthritis
predominantly affecting the spine and pelvis. It occurs with a
prevalence of 0.1\% to 1.4\% depending on the considered population
\cite{Sieper2002Ankylosing}. Genetic factors account for more
than 90\% of the risk of susceptibility to AS.
Human leukocyte antigen (HLA) class I molecule HLA
B27, belonging to the Major Histocompatibility Complex (MHC) region,
was the first genetic risk factor identified as associated with
ankylosing spondylitis in the 1970's \cite{Schlosstein1973High,
  Woodrow1978HLA} and remains the most important risk locus for this
pathology. Despite the strong association only a small portion of
HLA-B27 carriers develop the disease. Furthermore, studies in families
suggest that less than $50$\% of the overall genetic risk is due to
HLA-B27, which suggests that other genetic factors are involved \cite{Thomas2010Genetics}. A number of updated reviews on AS genetics, including genome-wide association study (GWAS) results,
identified new ankylosing spondylitis-susceptibility genes outside of the MHC region \cite{Tsui2014genetic, Reveille2010Genome}.

We applied all the methods described above to the  AS dataset. The data contain 408 cases
and 358 controls, and each individual was genotyped for 116, 513 SNPs with
Immunochip technology. For each SNP we obtained detailed genetic
information, such as 
gene affiliation, with the NCBI2R package
\cite{Melville2012Package} which annotates lists of SNPs with current
information from NCBI. We considered only SNPs located within a single
gene in order to form gene groups without overlap. We focused our analysis on a list of 29 genes previously identified as
having a main effect in GWAS.

The three methods tested yield different results, and only the PLS and G-GEE methods identify interactions. PCA detects only the main effect HLA-B and identifies no interactions. PLS detects the main effect HLA-B, but also identifies one interaction effect between the genes EOMES and BACH2.
Our method G-GEE does not detect any main effects, but it shows two significant interactions, the first between the genes HLA-B and SULT1A1 and the second between IL23R and ERAP2.

\subsubsection*{Thyroid carcinomas}
Thyroid cancers are thought to be related to a number of environmental and genetic predisposing factors and can be classified in various types and subtypes. Most association studies have focused on main effects but only a limited number of genes were identified. Recently, some papers focus on the detection of epistatic interactions \cite{Landa2013epistatic, Luzon-Toro2015Identification}. 
We applied our proposed approach on the two data sets used in Luz\'on-Toro el al. \cite{Luzon-Toro2015Identification} regarding two rare tumours, sporadic medullary thyroid carcinoma (sMTC) and juvenile papillary thyroid carcinoma (jPTC). Affymetrix Genome-Wide Human SNP 6.0 arrays were used to hybridized DNA. The data set related to sMTC contains 66 cases and the jPTC data set 30 cases. The same 125 healthy controls and 232, 607 SNPs were used for both studies.
As for the ankylosing spondylitis dataset, we obtained gene affiliation for each SNP with the NCBI2R package and considered only SNPs located within a single gene. We focused the analysis of the sMTC data set on a list of 10 genes, 3 of these genes (CHFR,  AC016582.2 and C8orf37) were chosen following the conclusions of Luz\'on-Toro el al., the others because they contained markers that were susceptible to be associated with the disease from univariate analysis. The analysis of the jPTC data set was realized on a list of 20 genes among them we can cite DIO, RP11-648K4.2, LOXL1, DMGDH, PAX8 and STK17B from  which epistatic interaction were already detected (even if  the interaction between PAX8 and STK17B was identified in a study concerning papillary thyroid and not the juvenile form). The 14 others genes contained susceptible associated individual markers from univariate analysis.
Regarding the sMTC study, G-GEE identifies one interaction between  genes NCK1 and TRIQK. PCA detects only one main effect for the gene TRIQK whereas none effects were identified with PLS. Concerning the jPTC data set, 3 interactions were identified by G-GEE  (NCAM1 and MNDA, MNDA and STK17B, LOC105370481 and STX3). PLS identifies 2 interactions (LOC105370236 and LOC105370481, LOC105370236 and PIKFYVE) and PCA detects only one main effect for the gene LOC105370481. We note that the effects detected with our approach concerned different genes that the ones identified in the presented previous studies (except for the gene STK17B). More analysese are needed to better understand these differences.

\subsubsection*{Inflammatory Bowel Disease}

Although the etiology of Inflammatory Bowel Disease (IBD) is not completely understood, previous studies have underlined the contribution of an important genetic susceptibility.  Recently, Martinez-Chamorro et al. \cite{Martinez-Chamorro2016Epistatic} detected an epistatic interaction between the genes NOD2 and TLR4. We applied our approach to the Wellcome Trust Case-Control Consortium genome-wide association dataset for Inflammatory Bowel Disease. The data contains 1949 case for 159 960 SNPs  genotyped by Affymetrix. The control group was constituted of 1972 individuals from the Wellcome Trust Case-Control Consortium genome-wide association dataset for hypertension. As for the two previous real data analysis, we obtained gene affiliation for each SNP with the NCBI2R package and considered only SNPs located within a single gene. The analysis was realized on a list of 22 genes that contain SNPs that are suspected to be associated with IBD from an univariate analysis. The two genes NOD2 and TLR4 were added to the list as they were previously detected as having an epistatic interaction. G-GEE identifies one interaction between the genes LOC105376008 and CACNB2 whereas PCA detects 9 main effects (IL23R, PODN, ATG16L1,
C5orf56, DNAH11, LOC105378282, HSD17B12, LINC00558, ADCY4) but none interaction. Finally PLS identifies 3 main effects for the genes IL23R, PODN and DNAH11 as well as 2 interactions the first one between the genes PODN and FCRLA, the second one between PVT1 and NOD2.

\section*{Discussion}

In this paper we compared different approaches for modelling gene-gene epistasis in a penalized regression framework. Our primary concern was the detection of interaction effects, and for this purpose we defined a general model and tested different interaction terms. We focused our analysis at the gene scale and compared three ways to design the interaction term.  Some methods were inspired by previous proposed approaches based on dimensional reduction methods including Principal Component Analysis (PCA) and Partial Least Square analysis (PLS). We additionally proposed a new interaction modeling approach that we called Gene-Gene Eigen-Epistasis (G-GEE),
where one interaction variable is built for each couple. The interaction variable was defined based on a criterion that maximizes the covariance between the phenotype and the pairwise SNP product matrix of the two genes.
The interaction components were then introduced in a Group Lasso penalized regression model that takes the gene structure into account and is capable of handling a large number of genes simultaneously.

A power study of the different methods based on two different simulation schemes (simplified and realistic) provided us with a rich body of information. Across various papers in the literature we find comparisons of similar methods that use different phenotype simulation settings. In the present work we compared two simulation models. Our first model was from a previous study \cite{Wang2014Pathway} that simulated the interaction component of each couple in an SNP pairwise product fashion. Our second model defined the interaction component as a pairwise product of representative variables of each gene.
Overall the G-GEE method performed well in detecting interactions in all the settings that were tested, although it was not always able to do so in the settings where main and interaction effects involved different genes. The power of the PCA method is highly dependent on the phenotype simulation model, because of the similarity between the second phenotype simulation model and the estimation model of the PCA method.
The PLS method is characterized by a lack of  power in detecting interactions. PLS performs well only when the related main effects are also present. When the simulated main and interaction effects do not concern the same genes, the detection capability of the PLS approach collapses dramatically.

For all methods we observed a confusion phenomenon when active genes are not simulated with both main and interaction effects. False detections of interactions concern genes that were simulated to have main effects, and false detections of main effects concern genes simulated to have interaction effects. This phenomenon reveals the difficulty that all methods encounter in clearly distinguishing the different types of effects. There are more false main detections when using methods such as PCA and PLS that are better at detecting main effects (except when the phenotype is simulated using the PCA model). As for interaction effects, the G-GEE methods make more false interaction detections than PCA and PLS. 

When genotypes are fully simulated in the simplified simulation study, the G-GEE and PCA approaches performed better when the PCA phenotype simulation model was used, whereas the PLS method was not very sensitive to the choice of phenotype simulation model. 
Unlike PCA and PLS, G-GEE is better at detecting interaction effects than at detecting main effects when simulations use a real data set. Since the simulation study using realistic data is meant to mimic real genotype data structure, we conclude that in a real context G-GEE will be better at detecting interaction effects than main effects.

In comparison to SNP-SNP interaction approaches, the gene-scale dimension of our proposed method means
that considerably fewer interaction variables need to be considered within a genetic region. This reduction in problem size allows larger problems to be handled. Moreover, a penalized regression method allows a true multivariate approach over a larger number of genes. It also extends other proposed gene-scale approaches, such as that presented by Wang et al. \cite{Wang2009partial}.
The ability to handle a relatively large number of genes simultaneously makes the detection of interactions between different genetic regions possible.
This might be useful as an initial step, prior to using SNP-SNP interaction methods that may provide more accurate information.

As the G-GEE method is not able yet to consider all human genes at the same time, it is necessary to specify a list of genes to be explored for potential interactions. Given that its power to detect main effects is low, for the detection of main effects it will be safer to use previously acquired  knowledge of the genetic effects, or to use a pre-processing method. 
Another limitation of the method  is gene size.  Computing the gene Eigen-Epistasis vector for two genes of size $p_r$ and $p_s$  requires an
$n \times (p_r p_s)$ matrix to be computed.

Prospects are improving the G-GEE method's performance by optimizing the computational cost and exploring new interaction functions to be plugged into the G-GEE criterion.

\paragraph{Availability of data and material:}

Our proposed G-GEE method has been implemented in an R package which is available on github: https://github.com/vstanislas/GGEE

\paragraph{List of abbreviations used}
\begin{itemize}[itemsep=-1ex]

\item 3G-SPA: model-based kernel machine method 
\item AS: Ankylosing Spondylitis
\item CCA: Canonical Correlation Analysis
\item CCU: Canonical Correlation-based U-statistic model 
\item GGEE: Gene-Gene Eigen-Epistasis
\item GWAS: Genome-Wide Association Studies
\item HLA: Human Leukocyte Antigen
\item LD:  Linkage Disequilibrium
\item MAF: Minor Allele Frequency
\item MHC: Major Histocompatibility Complex
\item PC:  Principal Components 
\item PCA: Principal Component Analysis 
\item PLS: Partial Least Squares 
\item SNP: Single-Nucleotide Polymorphism
\end{itemize}

\paragraph{Ethics (and consent to participate)}

The dataset regarding ankylosing spondylitis consists of the French subset of the large study of the International Genetics of Ankylosing Spondylitis (IGAS) study \cite{Cortes2013Identification}. For this subset, unrelated cases were recruited through the Rheumatology clinic of Ambroise Paré Hospital (Boulogne-Billancourt, France) or through the national self-help patients' association:  "Association Française des Spondylarthritiques". Population-matched unrelated controls were obtained from the "Centre d'Etude du Polymorphisme Humain", or were recruited as healthy spouses of cases. The protocol was reviewed and approved by the Ethics committee of the Ambroise Paré hospital. All participants gave their informed consent to the study. 

The application on thyroid carcinomas was carried out on a public dataset that came from the study of Luz\'on-Toro el al. on identification of epistatic interactions in two different types of thyroid carcinomas \cite{Luzon-Toro2015Identification}. This article is distributed under the terms of the Creative Commons Attribution 4.0 International License (http://creativecommons.org/licenses/by/4.0/), which permits unrestricted use, distribution, and reproduction.
All subjects underwent peripheral blood extraction for genomic DNA isolation using MagNA Pure LC system (Roche, Indianapolis, IN) according to the manufacturer's instructions.
A written informed consent was obtained from all the participants for clinical and molecular genetic studies. The study was approved by the Ethics Committee for clinical research in the University Hospital Virgen del Roc\'io (Seville, Spain) and complies with the tenets of the declaration of Helsinki.

Finally, the study on Inflammatory Bowel Diseases makes use of data generated by the Wellcome Trust Case-Control Consortium. A full list of the investigators who contributed to the generation of the data is available from www.wtccc.org.uk. 

\newpage

\bibliographystyle{vancouver}
\bibliography{StanBib}
\appendix

\subsection*{Additional Files}
\renewcommand\thefigure{S\arabic{figure}}  
\setcounter{figure}{0} 

\begin{figure}[!h]
\includegraphics[width=10cm]{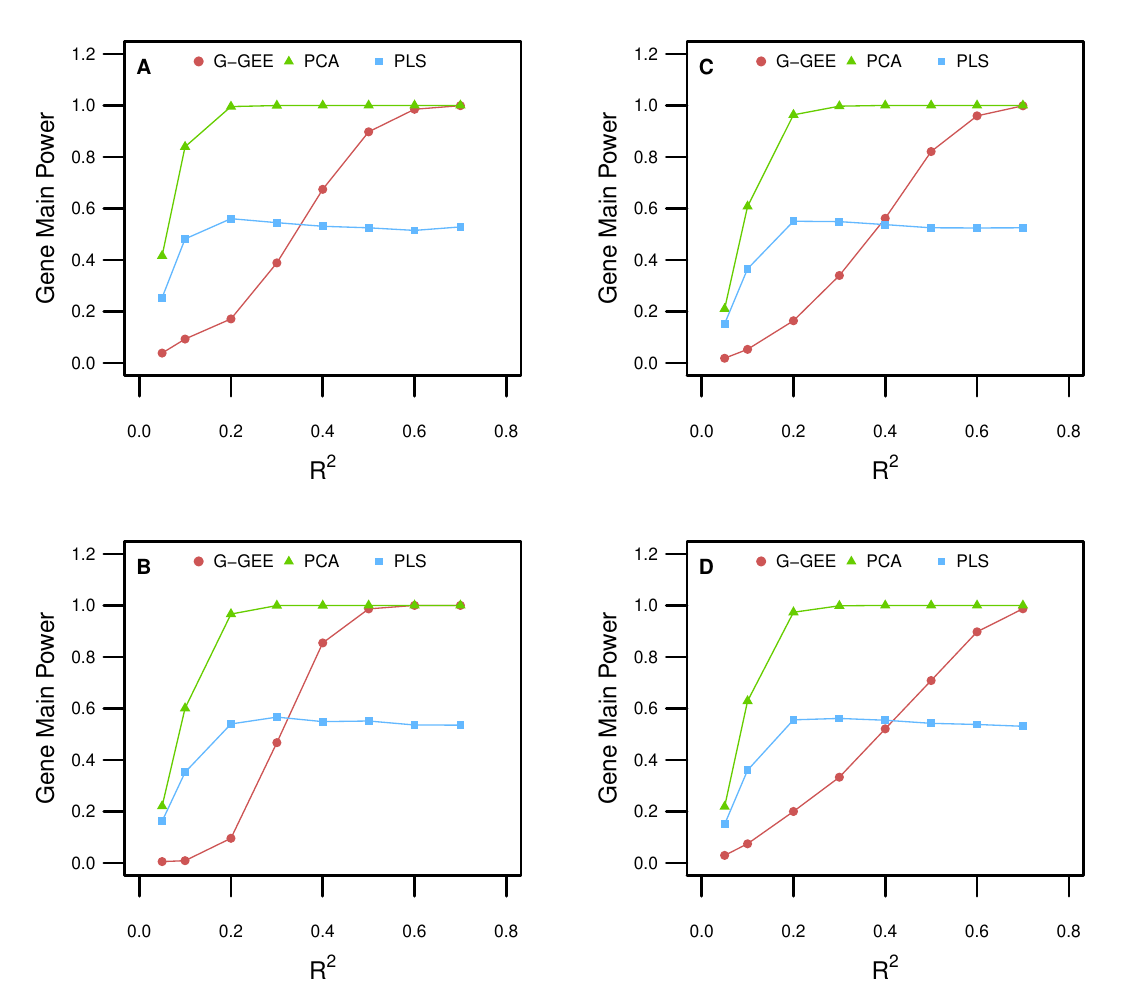}
            \caption{\textbf{Power under a simplified context}
             The figures show the power to detect main effects of the three methods depending on $R^2$.}
             \label{fig: CompaME1}
  \end{figure}

 \paragraph{Additional file 1 ---    Figure ~\ref{fig: CompaME1}}
  Comparison of power to detect main effects in the simplified simulation study.

\begin{figure}[!h]
\includegraphics[width=10cm]{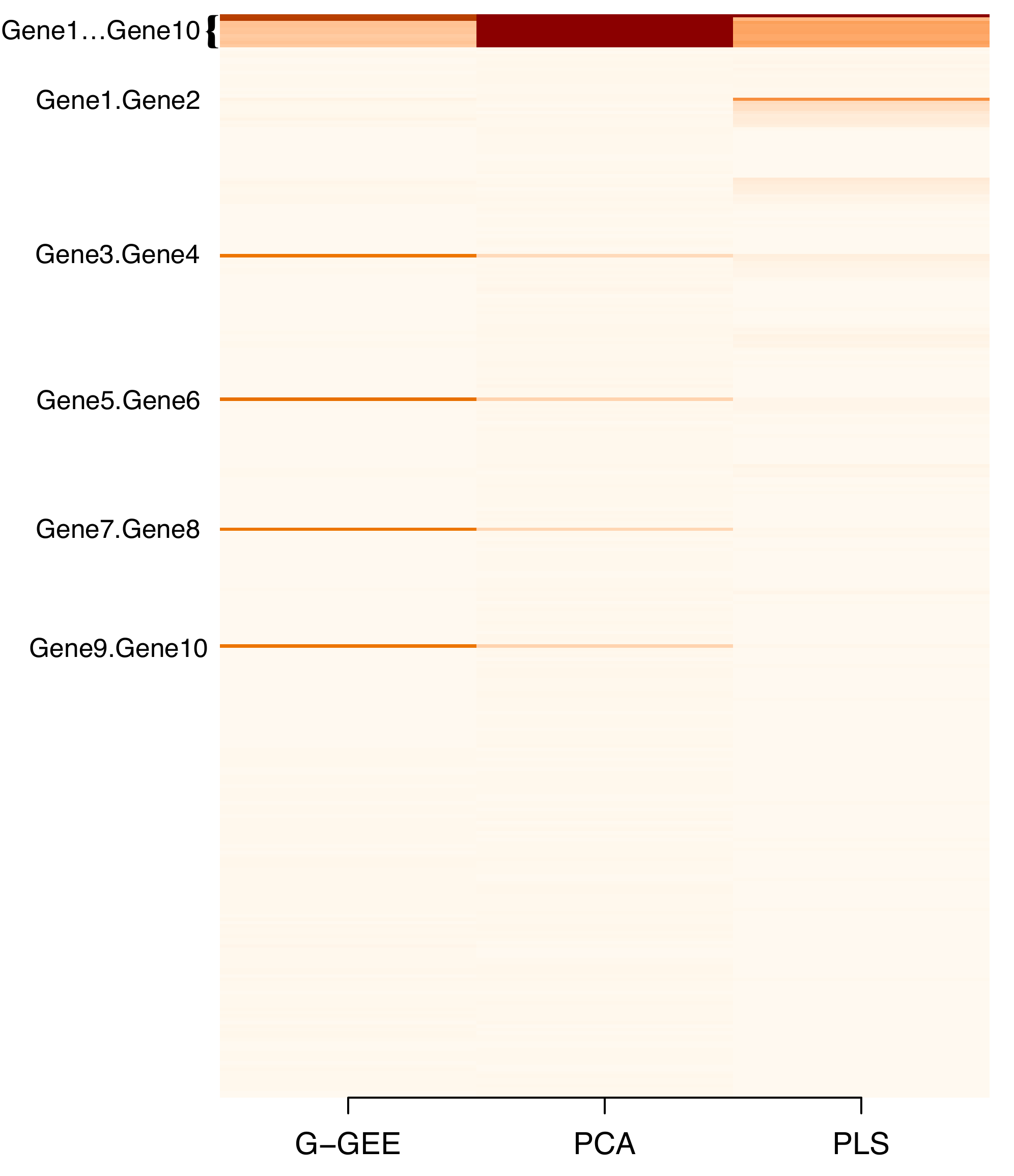}
        \caption{\textbf{Discoveries for the setting with numerous effects}
Heatmap of the ratio of the number of times where each variable was significant to the total number of simulations for $R^2=0.7$ using the Wang Pathway model for the phenotype simulation with fully simulated data. We consider 25 genes with two main effects for
genes 1 and 2, and four interaction effects between genes 3 and
4,  genes 5 and 6,  genes 7 and 8, and genes 9 and
10.  }
                 \label{fig: simu24}
\end{figure}

   \paragraph{Additional file 2 ---    Figure ~\ref{fig: simu24}}
Simulation on 25 genes with fully simulated data and various simulated effects.

\begin{figure}[!h]
\includegraphics[width=12.5cm]{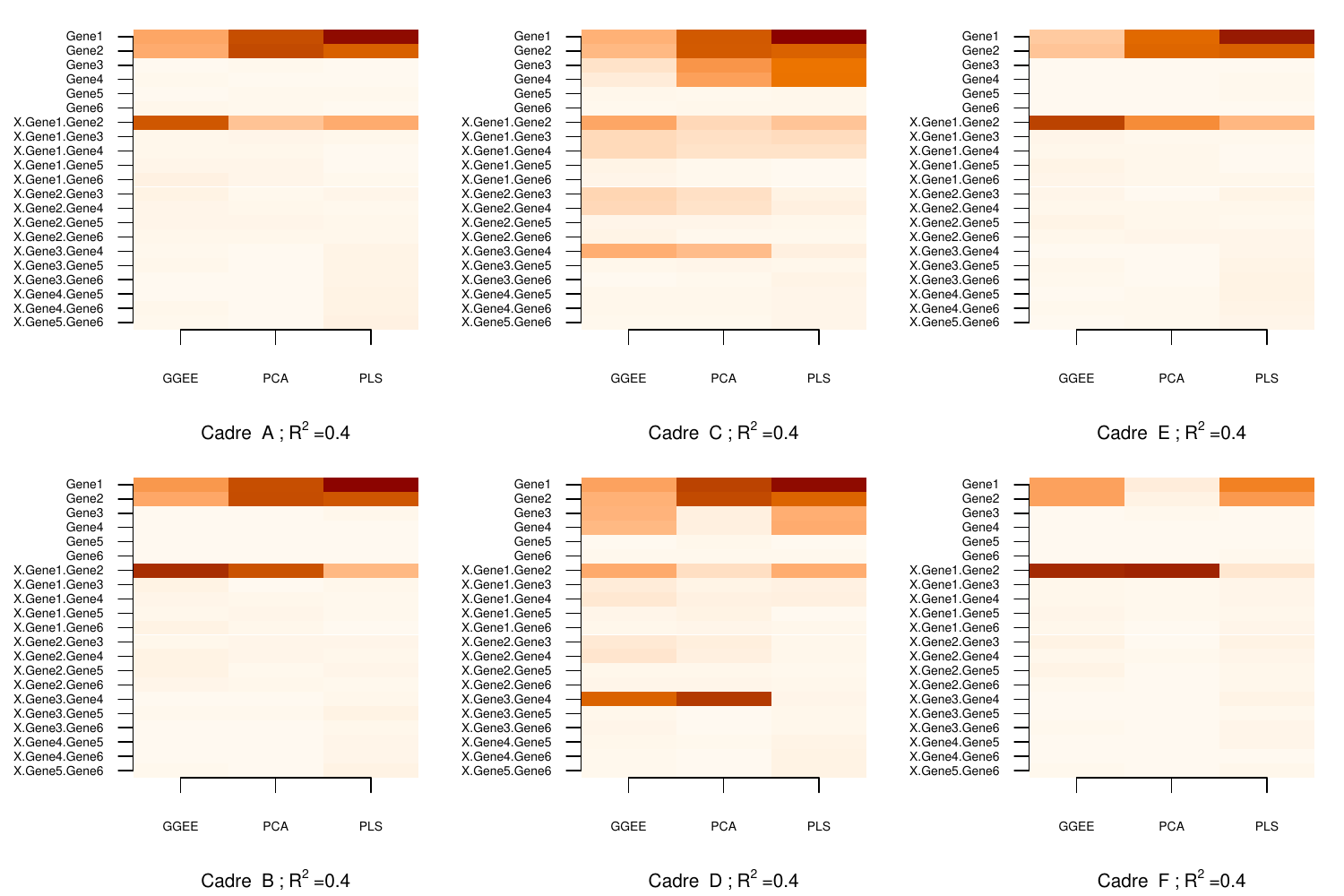}
        \caption{\textbf{Discoveries under a realistic context}
          Heatmaps of the ratio of the number of times where each variable was significant to the total number of simulations for $R^2=0.4$.}
               \label{fig: MATR4}
\end{figure}

 \paragraph{Additional file 3 ---   Figure ~\ref{fig: MATR4}}
Discoveries in the realistic simulation study with $R^2=0.4$.

\begin{figure}[!h]
\includegraphics[width=12.5cm]{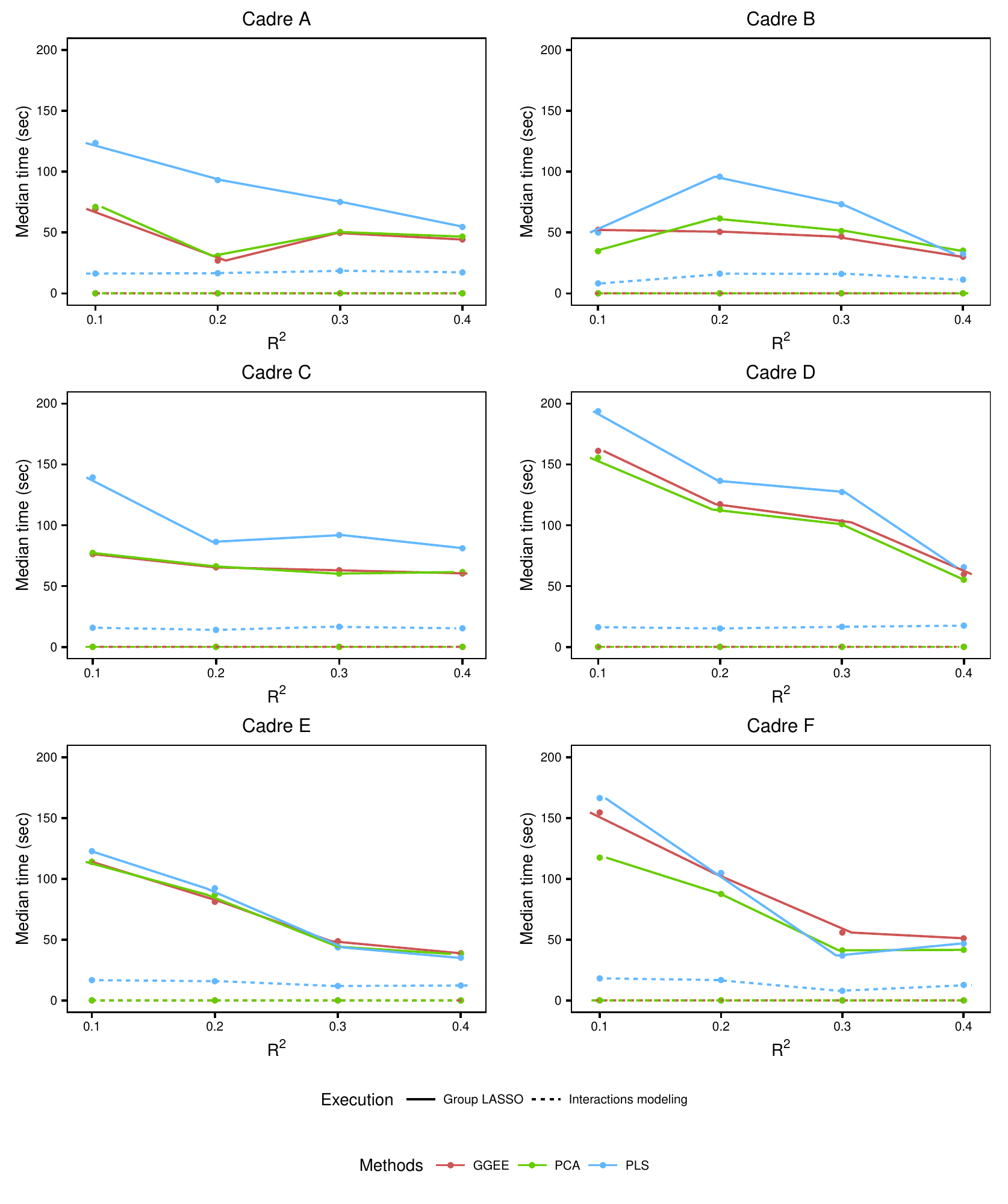}
        \caption{\textbf{Execution time}
      Median of the execution time to model interaction and to fit Group Lasso for the five first settings of the realistic simulation study.}
               \label{fig: ExecTime}
\end{figure}

 \paragraph{Additional file 4 ---   Figure ~\ref{fig: ExecTime}}
Comparison of execution time required to model interaction and to fit Group Lasso for the five first settings of the realistic simulation study.

\begin{figure}[!h]
\includegraphics[width=10cm]{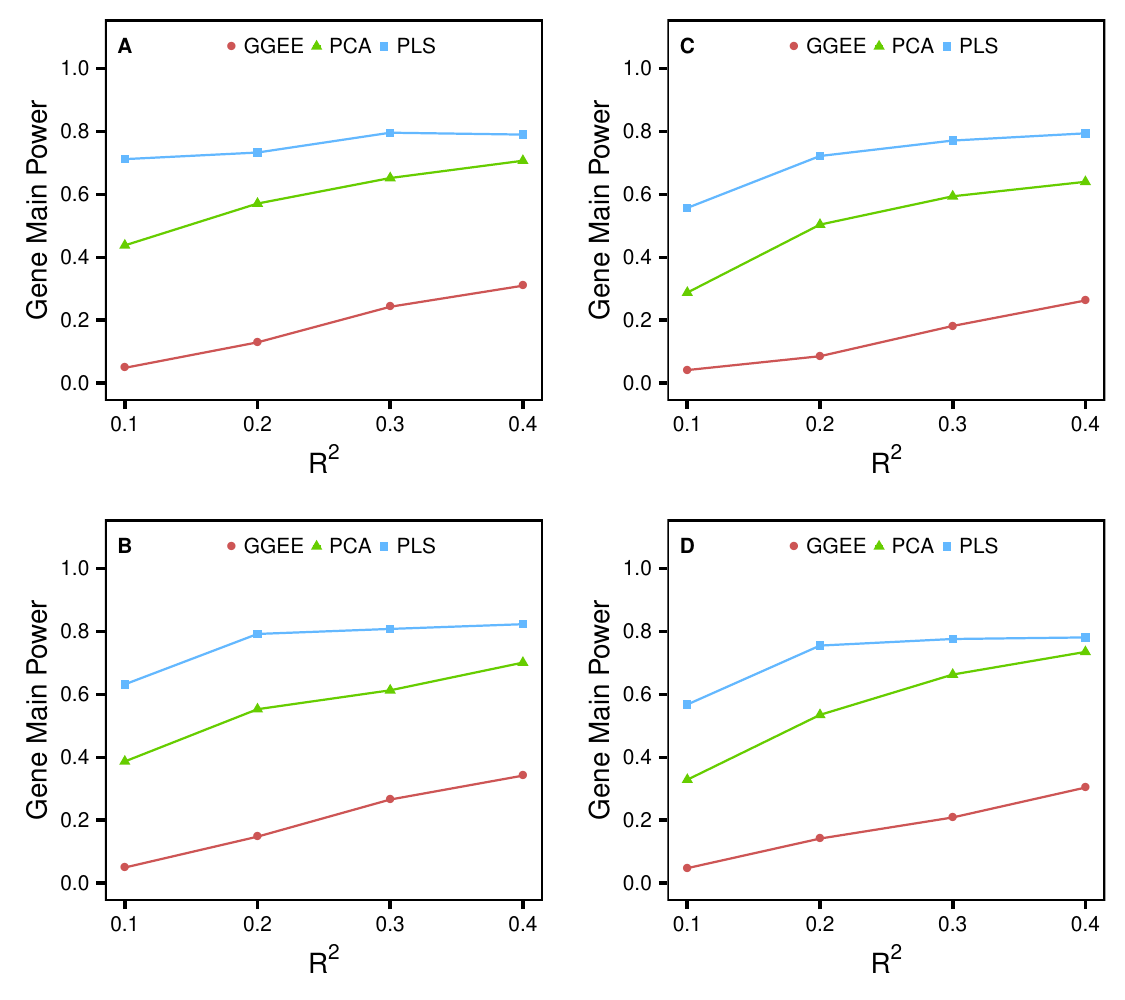}
            \caption{\textbf{Power under a realistic context}
             The figures show the power to detect main effects of the three methods depending on $R^2$.}
             \label{fig: CompaME}
  \end{figure}

 \paragraph{Additional file 5 ---   Figure ~\ref{fig: CompaME}}
 Comparison of power to detect main effects in the realistic simulation study.

\end{document}